**A quantitative system for discriminating induced pluripotent stem cells, embryonic stem cells, and somatic cells**

Anyou Wang[1*]

*Corresponding author

Anyou Wang

5109 Bioinformatics Building

University of North Carolina

Chapel Hill, NC 27599

*anyou.wang@alumni.ucr.edu*

**Running title:** Discriminating iPSCs, ESCs and SCs

**Abstract:** Embryonic stem cells (ESCs) and induced pluripotent stem cells (iPSCs) derived from somatic cells (SCs) provide promising resources for regenerative medicine and medical research, leading to a daily identification of new cell lines. However, an efficient system to discriminate the cell lines is lacking. Here, we developed a quantitative system to discriminate the three cell types, iPSCs, ESCs and SCs. The system contains DNA-methylation biomarkers and mathematical models, including an artificial neural network and support vector machines. All biomarkers were unbiasedly selected by calculating an eigengene score derived from analysis of genome-wide DNA methylations. With 30 biomarkers, or even with as few as 3 top biomarkers, this system can discriminate SCs from ESCs and iPSCs with almost 100% accuracy, and with approximately 100 biomarkers, the system can distinguish ESCs from iPSCs with an accuracy of 95%. This robust system performs precisely with raw data without normalization as well as with converted data in which the continuous methylation levels are accounted. Strikingly, this system can even accurately predict new samples generated from different microarray platforms and the next-generation sequencing. The subtypes of cells, such as female and male iPSCs and fetal and adult SCs, can also be discriminated with this system. Thus, this quantitative system works as a novel general and accurate framework for discriminating the three cell types, iPSCs, ESCs, and SCs and this strategy supports the notion that DNA-methylation generally varies among the three cell types.

**Introduction**

Embryonic stem cells (ESCs) and induced pluripotent stem cells (iPSCs) provide important resources for regenerative medicine and medical research [1,2,3,4,5]. Given the potential of these stem cell lines, the numbers of identified cell lines accumulate daily. In order to tap into this great resource, an accurate system to discriminate the cell lines is required. However, such a discriminant system remains to be developed.

Traditionally, biomarkers derived from well-characterized individual molecules have been used to distinguish somatic cells (SCs) versus pluripotent cells (PCs), including iPSCs and ESCs [6,7]. PCR and immunostaining can be used to further aid the biomarkers in distinguishing SCs from PCs [6]. However, applying the biomarkers to inherent multipotent cell lines could mislead the results due to the instabilities of multipotent cell lines that vary with conditions [7]. For examples, the OCT4 biomarker, which was once thought to be an excellent marker for discriminating ESCs and SCs, is only transitionally expressed in ESCs and is not consistently expressed in different ESCs, especially in old ESCs [7]. Any single biomarker that was selected from a very limited number of samples is unlikely to be robust enough to classify novel stem cells when applied alone across various conditions [7]. In addition, most of the current biomarkers based on antibodies will fail to detect the protein signals that are of low abundance, and thus the antibody-based biomarkers naturally exhibit low sensitivity. The strategy developed for distinguishing SCs from PCs also may not work for discriminating ESCs and iPSCs due to their similarity.

It is challenging to discriminate ESCs and iPSCs. Meta-analyses of genome-wide gene expression data sets which contain a large sample size can circumvent sample size limitations and generate the unbiased signatures needed to classify ESCs with the aid of cluster analysis [8]. A combination of linear models and gene expression profiling was also used to classify PCs and SCs [9]. However, the gene signatures cannot be used to distinguish iPSCs and ESCs because the gene signatures are not consistently expressed across different cell lines and conditions [10,11]. The gene expression profiling of iPSCs could be lab-specific when the batch effect was inappropriately adjusted [10,12]. Furthermore, liner models and clustering analysis are associated with a low sensitivity in determining classification, and they are also not the optimal mode of data classification with an abnormal distribution [13] and with different resources. Thus, the need for a system that can discriminate all three cell types remains.

In contrast to gene expression, DNA methylation consistently varies between iPSCs and ESCs under various conditions [14,15,16]. This suggests that signatures based on DNA methylation could be used as biomarkers to discriminate iPSCs and ESCs. In addition, SCs express distinct DNA methylation patterns compared to PCs [17]. Thus, DNA methylation-based biomarkers could provide a promising way to discriminate the three cell lines.

Applying mathematical models can quantitatively and sensitively discriminate biological samples [12,18,19]. If systems embedded with mathematical models are trained with large sample size, the system can also predict unknown samples. Among mathematical models, artificial neural network (NNET) [18] and support vector machines (SVM) [20] are frequently employed in biological discriminations [12]. NNET is a form of machine learning and non-linear statistical data modeling tool which processes data using a connectionism approach through an interconnected group of artificial neurons [18]. In most cases, NNET adjusts its structure during the learning phase according to external or internal information flowing through the network [18]. Therefore, NNET is able to cope with noisy and highly dimensional datasets, and the network has been utilized in many areas of medicine [18]. Similarly, SVM can discriminate complex samples as we previously reported [12].

In this study, we systematically selected biomarkers by an eigengene score [12], which was calculated from global methylation profiling, and we established a quantitative system to discriminate iPSCs, ESCs and SCs using mathematical models (i.e. NNET and SVM).

**Results**

**DNA methylation profiling of iPSCs, ESCs and SCs**

To investigate the DNA methylation profiling differentially expressed in iPSCs, ESCs and SCs, we analyzed genome-wide microarray profiling of these three cell types. In order to avoid cell-line-specific DNA methylation signatures and to develop a general system to discriminate the three cell types, we downloaded and analyzed a large set of data that contains as various sources as possible when they are relative to the three cell types (**Table S1,** materials and methods). A total of 636 microarrays were used in this study, including 55% SCs, 18% iPSCs, and 27% ESCs (**Table S1**). Various cell sources were included, such as male and female iPSCs, fetal and adult somatic cells, various tissues, fibroblasts, iPSC-derived and ESCs-derived somatic cells, fibroblast-derived iPSCs, ESCs-somatic cell-derived iPSCs, and epithelial cell-derived iPSCs.

Unsupervised cluster analysis and correspondence analysis of these samples revealed that

SCs are clearly separated from PCs, which include iPSCs and ESCs, and that most ESCs can be separated from iPSCs **(Figure 1A)**. While SCs are separated from PCs in first correspondence component, ESCs were somewhat different from iPSCs in respect to the correspondence component 2 and component 3 **(Figure 1B and 1C)**. This is consistent with the recent finding that iPSCs express distinct methylation profiling compared to ESCs [14] and suggests that the DNA methylation could be used as a variable to select biomarkers for distinguishing the three cell types.

**DNA methylation biomarker selection**

To improve the quantitative performance of our system, we selected biomarkers that contribute most of variances in this system. This ensures that selected biomarkers capture the primary features of data variations. To circumvent the correlations of gene methylations, we employed an eigengene ranking system derived from principal component analysis (PCA) as we previously reported [12] (materials and methods), instead of using traditional approaches that are mostly based on differential analysis. To be conservative and consistent, we only used filtered data from one abundant platform (illumina methylation 27K, GPL8490) to select biomarkers, and we used other platforms including next-generation sequencing data for testing (materials and methods).

We ranked all methylation loci by the eigengene score, and then selected the top ~200 methylation sites as biomarkers for each group, SCs versus PCs and iPSCs versus ESCs (**Table 1, Table S2-S3**). Interestingly, we found two groups of biomarkers for discriminating iPSCs and ESCs. Both groups are important in variance contributions, and they are distributed in two separate PCA components. Biologically, one group is located in autosomes and another in the x-chromosome (**Table S3-S4**).

**A quantitative system discriminating iPSCs, ESCs and SCs**

To establish a quantitative system for discriminating iPSCs, ESCs and SCs, we employed two types of mathematical models, artificial neural network (NNET) and support vector machines (SVM). We ran the above models with our data filtered by biomarkers. In both models, we measured the percentage of correction rate and kappa coefficient, which is a statistical measure of inter-rater agreement for quantitative items.

To determine the optimal biomarker number for the quantitative system of discriminating SCs from PCs, we ran both NNET and SVM by using a series of marker sets, which follow the order listed in table 1 and table S2. In each marker set, all samples were randomly sampled 200 times. In each run, 70% of random samples worked for training and the rest 30% for testing [12] (materials and methods). The accuracy of the 200 runs for each marker set was calculated.

With approximately 20 markers, both NNET and SVM discriminated SCs from PCs with an average percentage and kappa of 1.0 and 1.0, respectively (**Figure 2A**). Even with 3 markers (cg03273615, cg18201077, cg20217872), both SVM and NNET could successfully discriminate these two cell types with similar accuracy, with an average percentage and kappa of approximately 1.0 and 1.0, respectively (**Figure 2A**). This system became stable, even achieved a static state, after approximately 30 markers were applied. This stable state suggests that 30 markers might be good enough for discriminating any types of SCs and PCs.

Similarly, we also applied the above approach to discriminate ESCs from iPSCs using two group markers, an autosomal group and a x-chromosomal group (**Table S3-S4**). The autosomal group starts with a 0.75 percentage and 0.4 kappa in both NNET and SVM. A stable state is reached with a 0.95 percentage and 0.9 kappa with approximately 100 markers (**Figure 2B**). With 75 markers, the system reaches ~90% accuracy **(Figure 2B)**. X-chromosomal group begins with 0.6 percentage and 0.1 kappa and requires more than 300 markers to reach 87 percentage and 0.6 kappa value. It seemed that more biomarkers are required to reach higher accuracy and to achieve system stability (**Figure S1**). Our study indicated that the autosomal group performed better than x-chromosomal group and that SVM and NNET perform similarly in our biomarker sets. We thereafter used the autosomal group as the analysis in this study. This data indicated that discriminating iPSCs and ESCs requires at least 100 biomarkers. This also suggested that iPSCs and ESCs samples sources are very various, leading to the consequence that more biomarkers (>100) are required to make the system robust and stable. Together, our system, which includes math models (SVM and NNET) and DNA methylation markers, 30 biomarkers and 100 biomarkers respectively for distinguishing SCs from PCs and IPSCs from ESCs, can successfully discriminate three cell types, SCs, iPSCs, and ESCs.

**The system can be expended for general methylation measurement**

Methylation data measured by traditional experiments like bisulfite conversion counting are usually presented as discrete percentage. The discrete percentages are highly correlated with beta values that come from microarray data as evidenced by a high correlation that exists in beta values and percentage methylation levels measured by genome-wide bisulfite sequencing of ESCs [21](**Figure 3A**). To make our system more applicable to biological experiments, we converted the beta value to a discrete percentage level as listed in the following pairs, beta value/percentage >0.90/100, 0.9/90, 0.85/80, 0.8/70, 0.75/60, 0.7/40, 0.6/25, 0.55/20, 0.5/10, 0.4/5, 0.3/2, 0.17/1, and <0.17/0.

Using the conversion data, our system performs similar to the performance with unconverted data, and it reaches 100% and 90% accuracy with 30 markers and 100 markers respectively for discriminating SCs from PCs and iPSCs from ESCs (**Figure 3**). The high accuracy with converted and unconverted data suggests that our system can be used as a generalized application.

**Robustness and validation**

To further investigate the robustness of this system, we tested this system using raw data without global normalization (materials and methods). This system surprisingly works similarly to that with normalized data. With 30 markers and 75 markers respectively for discriminating SCs from PCs and ESCs from iPSCs, our system reaches 100% and 90% accuracy (**Figure 4**).

We validated this system using different platforms and resources including data from two new platforms, Illumna 450K (GPL13534) (materials and methods) and next-generation sequencing [16] and data generated by another research group looking at aging whose focus was unrelated to stem cell research [22]. The performance of our system was tested on each platform or resource. In the aging group [22], only SCs are available. All data was run with at least 20 sets of biomarkers; at least 30 to 50 markers were used for discriminating SCs from PCs and 170 to 200 markers were used for discriminating ESCs from iPSCs (materials and methods). This system can correctly predict SCs from PCs 100% of the time under all conditions, while it discriminates ESCs from iPSCs with ~90% of accuracy (**Table 2**). The accuracy rates suggest that this system is very robust and predictive.

**Cell subtype discrimination**

Distinguished DNA methylation patterns have been observed in subtypes of cells, such as the subtype of fetal and adult somatic cells and the subtype of female and male iPSCs [17]. Correspondence analysis of DNA methylation levels of SCs and iPSCs also showed that female iPSCs clearly separate from male iPSCs (Figure 5A) and adult SCs separate from fetal SCs (Figure 5B). This suggested that DNA methylation could be used to select biomarkers for discriminating cell subtypes. We used the same strategy described above to select DNA methylation biomarkers for discriminating two subtypes, iPSCs male and female subtypes and SCs fetal and adult subtypes (Table S5-S6). Mathematical models with these biomarkers showed that the accuracy of discriminating female iPSCs from male iPSCs reaches 100% accuracy when using 2 biomarkers (Table S5). When discriminating the adult SCs and fetal SCs, 95% accuracy is reached with 80 biomarkers. The high level of accuracy indicated that our system could be extended to identify the cell subtypes.

**Discussion**

For the first time, this study established a general quantitative system based on DNA-methylation markers to discriminate three cell types, iPSCs, ESCs, and SCs. Conventional methods like the OCT4 based method to distinguish ESCs from SCs have limitations and may not be efficient. Currently, there is no way to discriminate iPSCs from ESCs due to their similarity.

SCs are obviously different from PCs, which include iPSCs and ESCs. SCs exhibit DNA methylation patterns that can be distinguished from PCs in all observed conditions so far [17,23]. It is therefore reasonable to use DNA methylation as a variable to discriminate SCs from PCs although this system does not exist to date. In contrast, iPSCs closely mimic ESCs in many aspects such as colony morphology, even gene expressions and microRNA profiling [1,2,3,4,5,11,24]. Although previous studies showed that iPSCs generated from single cell resources display DNA methylation variations compared with ESCs [14,15,23], these variations could be cell-type specific and condition-dependent due to the limitation of its sample-size and the pure sample-resources. Here, we collected a large dataset including various cell line sources and conditions (Table S1) to determine if the DNA methylation pattern varied between iPSCs and ESCs (Figure 1). The DNA methylation pattern revealed that iPSCs generally exhibit certain variations compared to ESCs regardless of their originality and conditions. Therefore, DNA methylation could be used to select biomarkers for discriminating iPSCs from ESCs.

Biomarker selections should consider two major aspects, the generalizability of the sample and method efficiency. Condition-specific samples [25] like cell-line specific samples [7] could bias biomarker selections. To make our system generalizable, we minimized the cell-line bias selection and included various cell line sources (Table S1), such as different cell originality and gender. Methods based on differential values are usually employed to select biomarkers; however, these methods focus only on the significant differences between variables and fail to avoid variable correlations and redundant information from the multiple dimensional microarray data. Thus, these conventional approaches could harass biomarker selections

[26]. We selected the biomarkers by adopting the unbiased eigengene selection approach as we previously reported [12] (materials and methods). Eigengene-based selection takes care of the variable correlations and the redundant information of multiple variables, and it selects the independent elements that contribute to most of the variances in the entire dataset. All conditions like cell originality and other conditions have been taken into account and variations of conditions and cell-originalities have been reflected in the variance contributions. Thus, the selected biomarkers should be the most generalizable and the most important ones in this system. A quantitative system based on these selected biomarkers should perform better than that one based on biomarkers selected from differential comparisons. Indeed, while not reported here, we found that a system based on the differential methylation performed poorer than our system reported here in term of discriminant accuracy. Therefore, the way that we employed here to select biomarkers is efficient and the biomarkers selected from general data including various sources should be of general properties.

The sensitivity is of most concern for biomarker system development [7,25,27]. Conventionally, methods based on PCR or immunochemistry with a single biomarker like OCT4 have been frequently used in medical researches for distinguishing ESCs [6,7], but it is unlikely for these traditional approaches to provide a sensitive system to discriminate all cell types under all conditions given the substantial heterogeneity among the cell types [7]. Clustering analysis based on gene expression signatures was proposed to classify ESCs [8]; however its use is severely limited by the natural low accuracy associated with cluster analysis and the numerous signatures involved in the clustering. Ideally, a simple system should be developed, including a small panel of biomarkers that are easy measured and a mathematical model that quantitatively performs a sensitive judgment. However, it is challenging to assemble and validate such a biomarker panel. Here, we employed a machine learning system based on NNET and SVM to systematically and quantitatively validate a panel of ~200 biomarkers for each group **(Figure 2).** NNET and SVM, combined with dimension-reduced approaches like principal component analysis, are advantageous when handling non-linear functions for nosey multiple dimension data and have been successfully applied in discriminating disease cell lines and molecular complexity [12,18]. NNET and SVM with as few as 3 biomarkers for determining SCs from PCs and with 100 markers for determining iPSCs from ESCs can discriminate the three cell types with 100% and 95% accuracy respectively for two groups (**Figure 2**). This suggests that our system is the most sensitive system to discriminate the three cell types to date.

Robustness and prediction value are also of concern in developing discriminant system [7,25,27]. Conventional approaches such as PCR, immunostaining and clustering analysis are of low robustness and prediction value. We tested our system with raw data without global normalization and with discrete methylation percentage data converted from continuous variables measured from microarray, and we found that our system 100% and 90% correctly discriminates SCs from PCs and iPSCs from ESCs, respectively (**Figure 3 and Figure 4**). When validated by new samples generated from other independent groups and even from different microarray platform and next-generation high throughput sequencing data, our system continued to correctly predict 100% SCs and 90% of iPSCs from ESCs (**Table 2**). Thus, this system established here is very robust and can be generally applied to discriminate the three cells types in medical research.

Furthermore, Nazor et al recently revealed the distinguished DNA methylation patterns existing in the subtypes of cells [17], such as the subtype of female against male in iPSCs, and the subtype of fetal versus adult in SCs. This suggested that DNA methylation could be used as a variable to discriminate the subtypes of cells. We extended our system to discriminate the subtypes of cells, and our system reached 100% accuracy in discriminating female and male iPSCs and 95% accuracy in classifying adult and fetal SCs (Figure 5). Concerning the SCs, which contained various tissues with tissue-specific methylation loci, the 95% accuracy with 80 biomarkers, would be very promising, and this system could be extended to discriminate other subtypes of cells when more data is available.

The methylation data for our biomarkers can be measured using traditional methods, and the measurement is less expensive than microarray and antibody-based immunochemical approach. Therefore, this system developed here is a cost effective, accurate and reliable discriminant system to distinguish three cell types. This approach lays a fundamental strategy to develop other discriminant systems.

**Materials and Methods**

**DNA methylation data and processing**

All 636 methylation microarray data were downloaded from GEO database (www.ncbi.nlm.nih.gov/geo/) (Table S1). The methylation data was preprocessed by GenomeStudio (http://www.illumina.com/gsp/genomestudio_software.ilmn) and then was processed using R (http://www.r-project.org/). All methylation values measured by microarray were calculated as beta value, ranging from 0 to 1. These values were pre-normalized by GenomeStudio using default parameters. Before further analysis, outliers were filtered out in basis of QC checking the distributions of X-chromosome beta value [28] and CpG methylation [29] and the euclidean distance from samples to group center. After outliers were filtered, only 312 out of 399 microarrays generated from the platform GPL8490 (Illumina 27k) were available for biomarker selection. Because the microarray data were not generated at the same time, the batch effect needs to be filtered out before combining the microarray datasets. An algorithm called ComBat [30], which runs in R environment and uses parametric and nonparametric empirical Bayes frameworks to adjust microarray data for batch effects, was used to adjust the final methylation values for all datasets.

**Biomarker selection**

Biomarkers were selected based on an eigengene score, which was defined below as we previously reported [12].

score = $|cor(x_i,E)|$

$|Cor(x_i, E)|$ is the absolute value of Pearson correlation coefficient, where $x_i$ is a vector of methylation of $i^{th}$ node, and E eigenvalue derived from principal component analysis.

**Artificial neural networks and support vector machines**

Mathematical models from NNET packages in R were used to perform artificial neural network (NNET). NNET is machine learning mathematical model that is designed to emulate the architecture of the brain[31]. In NNET, data is processed by neurons that are organized in parallel layers: input, hidden, and output. The neurons of the input layer receive data as a methylation value and transmit the input data into the hidden layer neurons that process the data using mathematical functions. The processed results are displayed into the output layer neurons. The output neuron with largest value in output layer will be the group that input neuron (either iPSCs, ESCs, or SCs) should be.

We used SVM as we previously reported [12]. Briefly, SVM classifies datasets based on hyperplanes in which samples can be clustered with the largest separated distances. The R package e1071 was used in this study.

For both NNET and SVM, we randomly sampled 200 times for each biomarker set, from 1 biomarker to 200 biomarkers, and we used 70% of the samples as a training set and the remaining 30% as test data. The accuracy was calculated from the test data set by measuring both average percentage correct rate and kappa value.

**Validation and prediction**

During validation and prediction, data were normalized from 0 to 1 number in basis of beta value format from Illumina Inc, without removing batch effect and without further normalization. All 27k platform data used for biomarker selections was treated as training data, and the samples from 45k platform (GPL13534), the next-generation sequencing [16], and the samples from the aging study [22] were used as separate testing sets. The testing samples were randomly sampled 200 times, each time using 90% of the samples as input for calculating the accuracy. At least 20 biomarker sets were run for each testing set, utilizing 30 to 50 markers to determine SCs from PCs and 170 to 200 markers to determine ESCs from iPSCs. Only markers that overlapped between training and testing data sets were used for each run. In the sequencing data, we used the read counts from methylation sites generated from sequencing data that was further standardized between 0 and 1.

**Acknowledgement**

We specially thank Dr. Amanda Jackson for her critically editing this paper. Not competing Interest exists in this paper.

**Figure and Table legends**

**Figure 1. Overall methylation profiling of three cell types.** A, Clustering analysis revealed that SCs were separated from PCs (iPSCs and ESCs). In the PCs subgroup, most ESCs were separated from iPSCs. B, Correspondence analysis classified three cell types, SCs, iPSCs, and ESCs. SCs and PCs were separated in first component while most of ESCs and iPSCs were separated in second component. C, iPSCs and ESCs were further classified in 3D space. For visualization purposes, only one subset of data was shown here.

**Figure 2. Performance of DNA methylation biomarker system.** Accuracy was measured as kappa value and accuracy percentage, shown on the Y-axis. The top panel A represents SCs discriminating from PCs and the bottom panel B represents ESCs from iPSCs. The X-axis represents marker number, from 1 marker to 50 markers in SCs versus PCs panel (top), and from 1 to 200 markers in ESCs versus iPSCs panel (bottom B). Only data for 50 and 200 markers for these two groups are shown because the system became a static state after that level.

**Figure 3. The discriminating system performs precisely on converted data.** A, a high correction relationship exists between methylation percentage measured from sequencing and the beta value measured from Illumina microarray. B, Our system discriminates the three cell types with high accuracy with converted data. For visualization purposes, only percentage of NNET was shown here due to its similarity with SVM and the high correlation between accuracy percentage and kappa. This practice was also applied to following figures in this study.

**Figure 4. Our system works accurately with raw data.** Our system reaches the similar discriminating power as that with normalized data.

**Figure 5. Cell subtype discrimination.** A and B denote correspondence analysis to classify subtypes of cells. A, the subtype of female and male iPSCs. B, the subtype of fetal and adult SCs. C and D show the accuracy of discriminating subtypes of cells. C, the subtype of female and male iPSCs. D, the subtype of fetal and adult SCs.

**Table 1, Top biomarker list.** Left panel, top 10 biomarkers for discriminating SCs from PCs. Right panel, top 10 biomarkers for discriminating iPSCs from ESCs. Please see Table S2-S4 for complete list used in this study.

**Table 2. Prediction profiling of our system.**

Figure 1

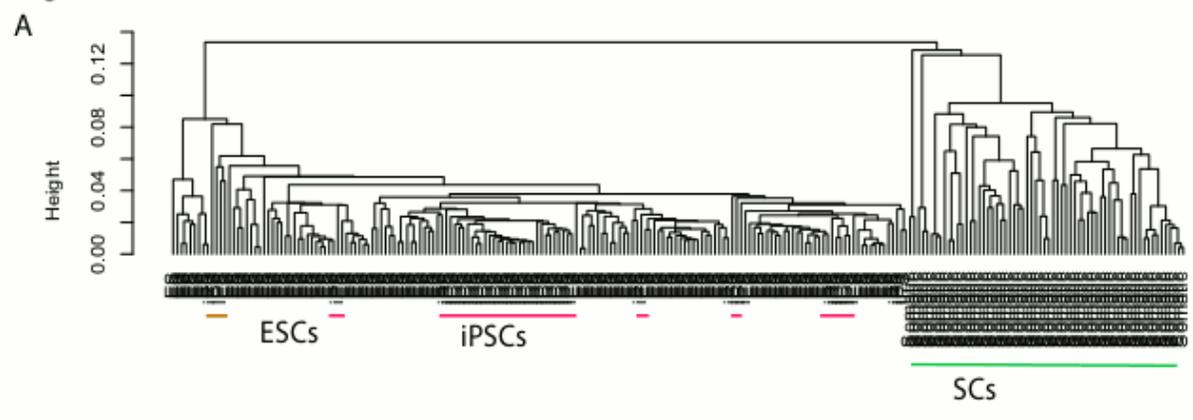

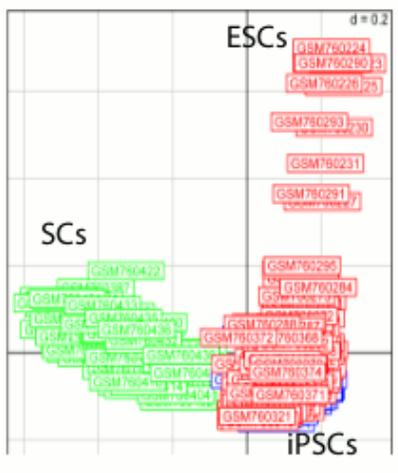
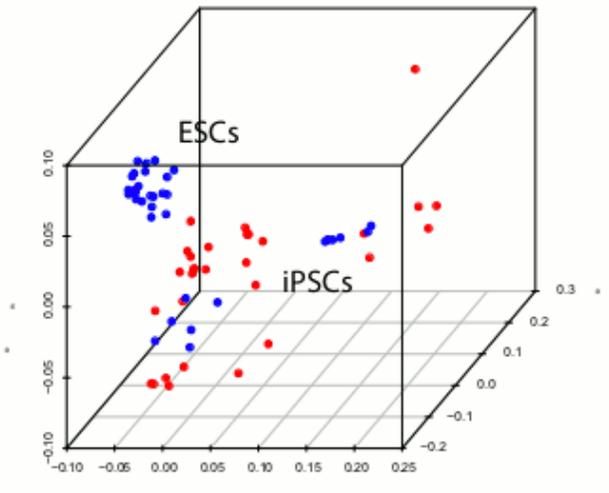

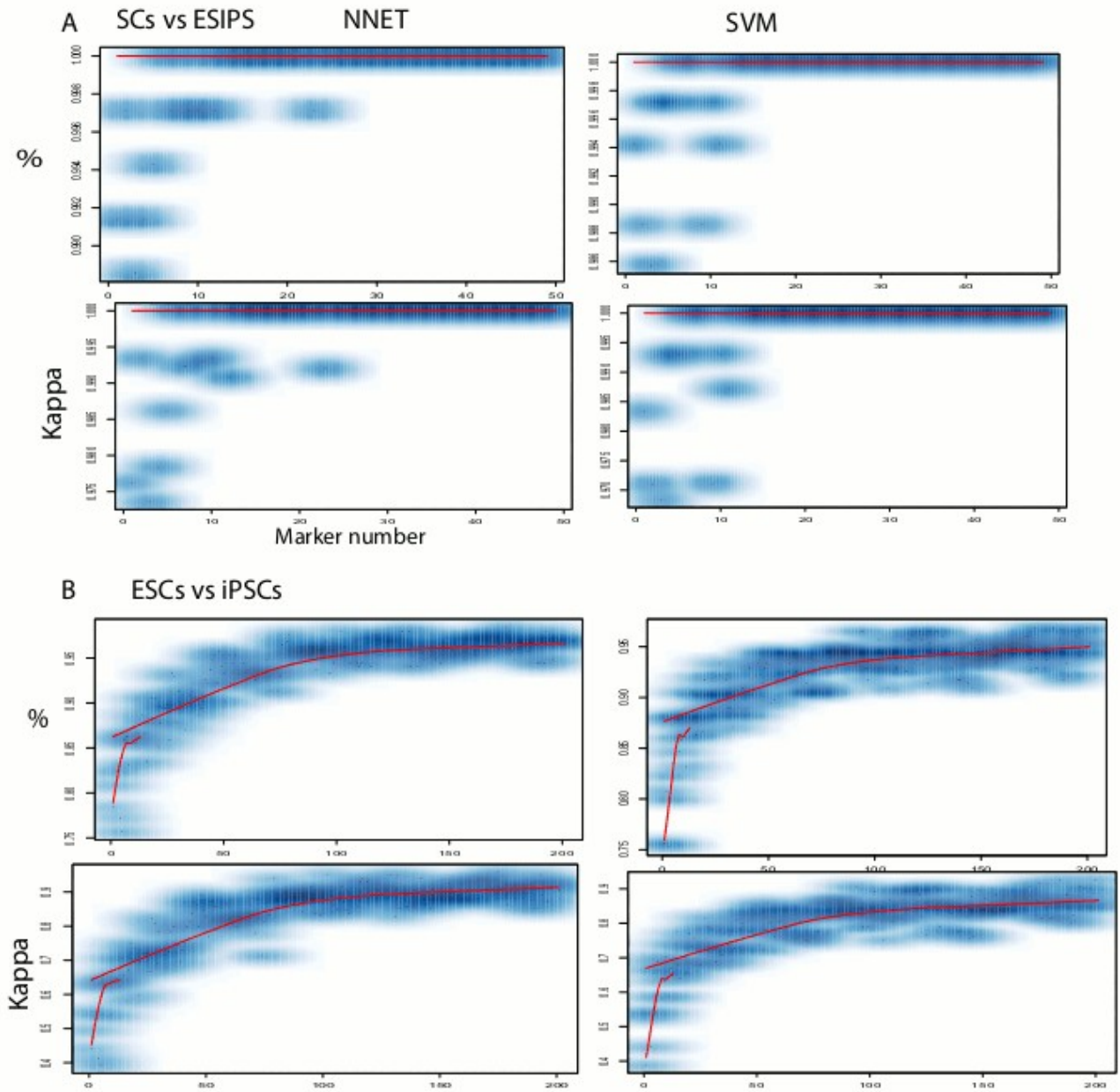

Figure 2

Figure 3

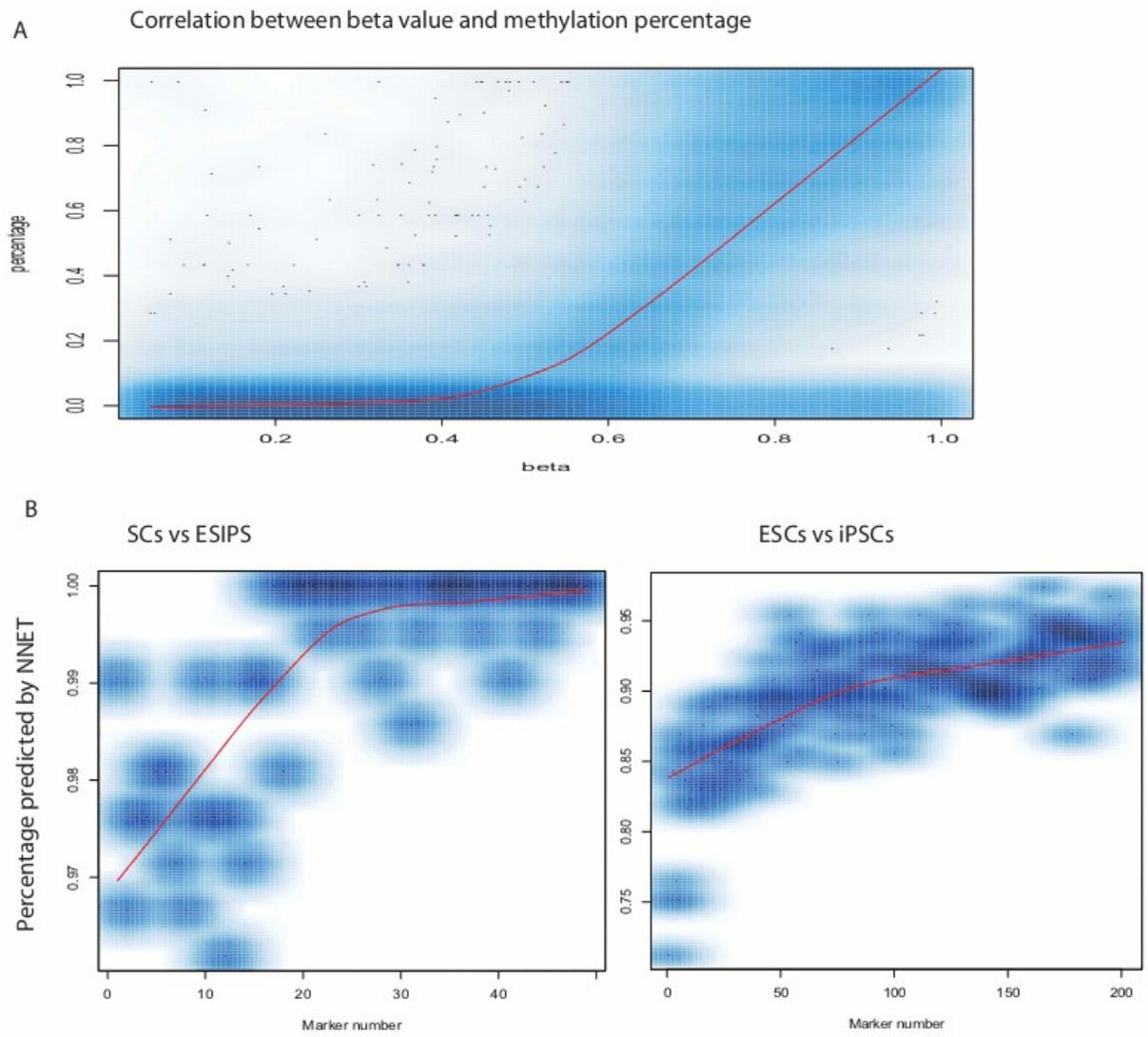

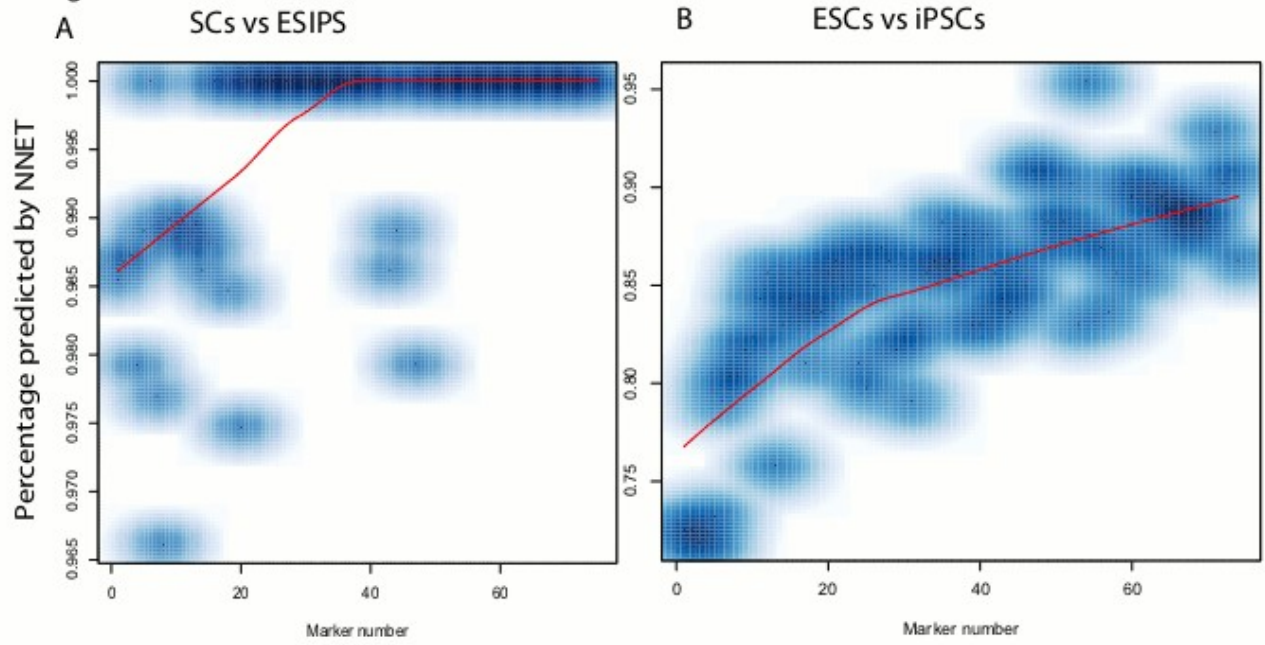

Table 1

### A  Top biomarkers for SCs vs ESIPS

| ID | Chr | MapInfo | Symbol | Ranking |
|---|---|---|---|---|
| cg03273615 | X | 106249029 | FLJ11016 | 0.85761406 |
| cg18201077 | 2 | 6935238 | RSAD2 | 0.85699794 |
| cg20217872 | 12 | 76748611 | NAV3 | 0.8556891 |
| cg25193278 | 6 | 26548763 | BTN3A3 | 0.8537096 |
| cg01337047 | 18 | 27151111 | DSG1 | 0.85273852 |
| cg11009736 | 2 | 119416152 | MARCO | 0.85190368 |
| cg05360220 | 1 | 2486381 | TNFRSF14 | 0.84817272 |
| cg02332073 | 7 | 130022959 | TSGA13 | 0.8467108 |
| cg04000821 | 19 | 59705878 | LAIR2 | 0.84633995 |
| cg03791917 | X | 100527943 | BTK | 0.84564686 |

### B  Top biomarkers for ESCs vs iPSCs

| ID | Chr | MapInfo | Symbol | Ranking |
|---|---|---|---|---|
| cg09527362 | 6 | 74218863 | C6orf150 | 0.891412 |
| cg22736354 | 6 | 18230698 | NHLRC1 | 0.88084 |
| cg19005368 | 11 | 32808526 | PRRG4 | 0.877542 |
| cg13628514 | 12 | 1.09E+08 | TRPV4 | 0.872649 |
| cg08946332 | 17 | 6840612 | ALOX12 | 0.871802 |
| cg03699904 | 3 | 1.72E+08 | SLC2A2 | 0.871177 |
| cg20019546 | 7 | 37922349 | SFRP4 | 0.870126 |
| cg00463577 | 6 | 74218632 | C6orf150 | 0.868448 |
| cg00815605 | 14 | 73105635 | ACOT2 | 0.868257 |
| cg21233722 | 5 | 1.69E+08 | DOCK2 | 0.867968 |

Table 2

| Quantile | 450K | | Sequencing | | aging |
|---|---|---|---|---|---|
| | SCs vs PCs | ESCs vs iPSCs | SCs vs PCs | ESCs vs iPSCs | SCs vs PCs |
| 0% | 1 | 0.85 | 1 | 0.85 | 1 |
| 25% | 1 | 0.87 | 1 | 0.88 | 1 |
| 50% | 1 | 0.88 | 1 | 0.9 | 1 |
| 75% | 1 | 0.89 | 1 | 0.92 | 1 |
| 100% | 1 | 0.91 | 1 | 0.93 | 1 |

Figure 5

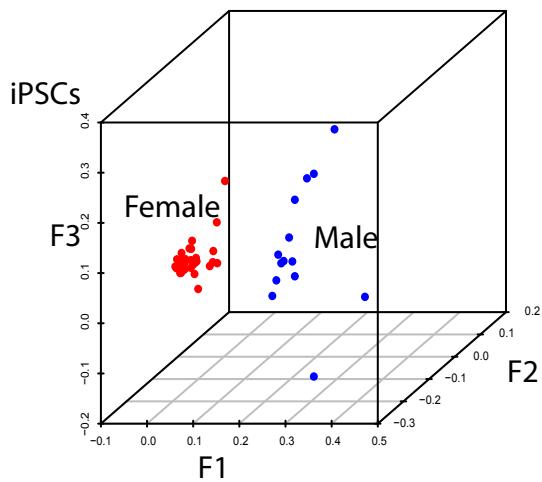

A — Classification — iPSCs

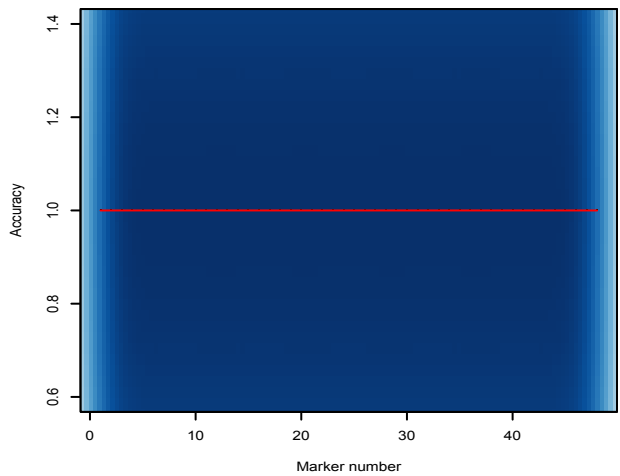

C — SVM

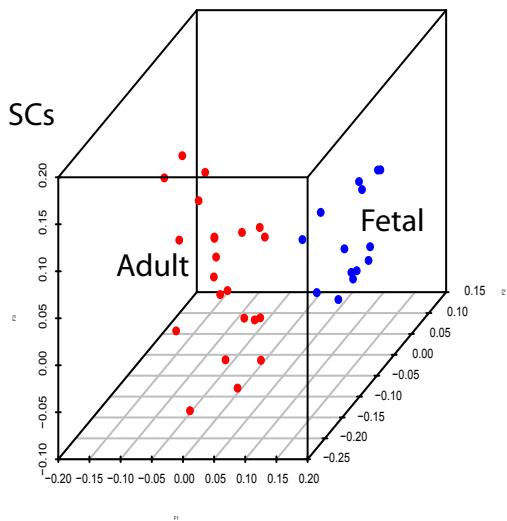

B — SCs

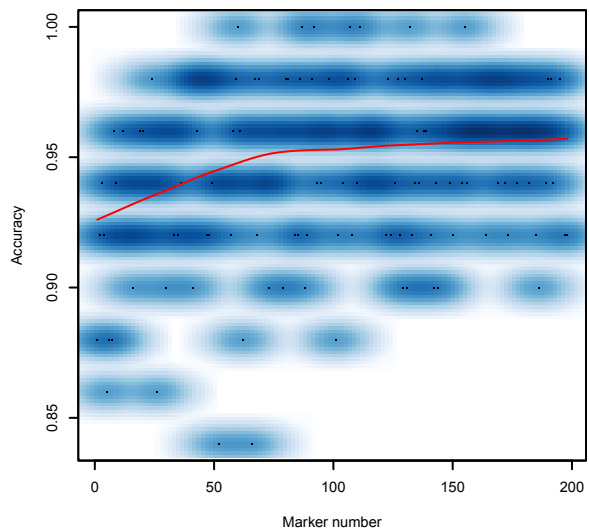

D